\documentclass{aastex62}
\usepackage{amsmath,amssymb,CJK}
\usepackage{gensymb}
\usepackage{soul}
\usepackage{ulem}
\usepackage{threeparttable}

\graphicspath{{./}{figures/}}

\submitjournal{ApJL}

\shorttitle{Water production rates and activity of interstellar comet  2I/Borisov}
\shortauthors{Xing et al.}

\begin{document}
\begin{CJK*}{UTF8}{gbsn}

\title{Water production rates and activity of interstellar comet 2I/Borisov}
\correspondingauthor{Dennis Bodewits}

\author[0000-0003-2399-5613]{Zexi Xing (邢泽曦)}
\affiliation{Department of Physics and Laboratory for Space Research, The University of Hong Kong, Pokfulam Road, Hong Kong SAR, China}
\affiliation{Physics Department, Leach Science Center, Auburn University, Auburn, AL 36849, USA}
\email{zexixing@hku.hk}

\author[0000-0002-2668-7248]{Dennis Bodewits}
\affiliation{Physics Department, Leach Science Center, Auburn University, Auburn, AL 36849, USA}
\email{dennis@auburn.edu}

\author[0000-0003-2152-6987]{John Noonan}
\affiliation{Department of Planetary Sciences, Lunar and Planetary Laboratory, University of Arizona, Tucson, AZ 85721, USA}

\author[0000-0003-3257-4490]{Michele T. Bannister}
\affiliation{Astrophysics Research Centre, School of Mathematics and Physics, Queen's University Belfast, Belfast BT7 1NN, United Kingdom}
\affiliation{School of Physical and Chemical Sciences --- Te Kura Mat\={u}, University of Canterbury,
Private Bag 4800, Christchurch 8140,
New Zealand}



\begin{abstract}
We observed the interstellar comet 2I/Borisov using the Neil Gehrels-\emph{Swift} Observatory's Ultraviolet/Optical Telescope. We obtained images of the OH gas and dust surrounding the nucleus at six epochs spaced before and after perihelion ($-2.56$~AU to 2.54~AU). Water production rates increased steadily before perihelion from $(7.0\pm1.5)\times10^{26}$ molecules~s$^{-1}$ on Nov.~1, 2019 to $(10.7\pm1.2)\times10^{26}$ molecules~s$^{-1}$ on Dec.~1. This rate of increase in water production rate is quicker than that of most dynamically new comets and at the slower end of the wide range of Jupiter-family comets. After perihelion, the water production rate decreased to $(4.9\pm0.9)\times10^{26}$ molecules~s$^{-1}$ on Dec.~21, which is much more rapidly than that of all previously observed comets. Our sublimation model constrains the minimum radius of the nucleus to 0.37~km, and indicates an active fraction of at least 55\% of the surface. $A(0)f\rho$ calculations show a variation between 57.5 and 105.6~cm with a slight trend peaking before the perihelion, lower than previous and concurrent published values. The observations confirm that 2I/Borisov is carbon-chain depleted and enriched in NH$_2$ relative to water. 
\end{abstract}




\section{Introduction}
\label{sec:intro}

Comets contain abundant amounts of organic and inorganic species \citep{Alt18} and observations of their chemical composition play an important role in reconstructing the conditions of planet formation in our solar system (e.g. \citealt{Ahe12}). Extrasolar comets offer a glimpse into the building blocks, formation, and evolution of other planetary systems, but only a limited number of gas species that can be directly attributed to the presence of exocomets have been detected to date \citep{Kra16,Mat17}. Impacts by exocomets may significantly alter the atmospheres of exoplanets \citep{Kra18} or introduce species that could be conceived as biosignatures \citep{Sea16}. Interstellar comets might also furnish the exchange of volatiles and complex molecules between different planetary systems. The existence of interstellar interlopers in our solar system has long been suggested \citep{Lev10, Ray18}, but their based on chemical signatures has been hampered by the lack of understanding of what drives the chemical taxonomy of comets in the first place \citep{Sch08,Del16}. With an eccentricity of 3.357, there is no question regarding the extrasolar origins of 2I/Borisov (Minor Planet Center; MPEC 2019-W50). At the time of its discovery, at 3~AU from the Sun, Borisov featured a prominent tail and coma. This implies that it contains sublimating volatiles, first confirmed by the detection of the emission of gaseous CN \citep{Fit19}. However, sublimation of H$_{2}$O is far more prolific in previously observed Solar System comets and thus provides a standardized tool to compare evolutionary, if not primordial, chemical abundances between comets. Here, we report on our 5-month long monitoring campaign  with the Neil Gehrels-\emph{Swift} observatory before and after its perihelion on Dec. 8.55 UTC to determine the water production rate of 2I/Borisov.

\section{Observations} \label{sec:obs}

The Neil Gehrels-\emph{Swift} Observatory \citep{Geh04} is a multi-wavelength satellite, originally designed for the rapid follow-up of gamma-ray bursts. In this study we used its UltraViolet Optical Telescope \citep[UVOT;][]{Rom00} to determine water production rates and the dust content of 2I/Borisov. The seven broadband filters of UVOT cover a range of 160--800 nm. UVOT has a 17~$\times$~17 arcmin field of view with a plate scale of 1 arcsec/pixel, and a point spread function (PSF) of approximately 2\farcs4  FWHM.

\emph{Swift}/UVOT observed 2I/Borisov using the V (central wavelength 546.8~nm, FWHM 76.9~nm) and UVW1 (central wavelength 260 nm, FWHM 69.3 nm) filters six times between 2019~September~27 and 2020~February~17~UTC (Table~\ref{tab:obs-log}). To minimize smearing caused by the comet's apparent motion ($\approx$ 3 -- 7 pix/200~seconds), each of the first five observations consists of 16 and 48 exposures of about 200~seconds respectively for the V filter and the UVW1 filter, whereas 4 V-band exposures were lost  for the January observation. Because the comet became much fainter and Swift was near its pole constraint, the final observation consists of 28 V-band exposures of about 200 or 85 seconds, and 98 UVW1-band exposures of about 200 seconds. Orbital information in Table~\ref{tab:obs-log} is from JPL Small-Body Database\footnote{JPL Small-Body Database: \url{https://ssd.jpl.nasa.gov/sbdb.cgi}}.

\begin{table}[h]
\caption{Summary of the observing log}
\label{tab:obs-log}
\centering
\scalebox{1}{
\hspace*{-5mm}
\makebox[\textwidth]{%
\tabcolsep=0.5cm
\scriptsize{
\begin{tabular}{*{10}c}
\hline
\hline
\# & Mid Time& T-T$_\mathrm{p}$& $r_\mathrm{h}$& $\Dot{r}_\mathrm{h}$& $\Delta$& S-T-O& UVW1 $T_\mathrm{exp}$& V $T_\mathrm{exp}$\\
&(UT)& (days) & (AU)& (km\,s$^{-1}$)& (AU)& (\degree)& (s)& (s) \\
\hline
1 & 2019-09-27\,08:52:40 & -72.2 & 2.56& -23.54& 3.10& 17.31& 8205 (8205)& 3099 (2712) \\
2 & 2019-11-01\,19:52:26 & -36.7 & 2.17 & -14.43& 2.42& 24.24& 7203 (5487)& 3098 (1935) \\
3 & 2019-12-01\,12:17:04 & -7.0 & 2.01 & -3.00& 2.04& 28.12& 8147 (5071)& 3092 (386) \\
4 & 2019-12-21\,12:52:31 & 13.0 & 2.03& 5.48& 1.94& 28.60& 8199 (6346)& 3100 (1937) \\
5 & 2020-01-14\,10:36:11 & 36.9 & 2.17& 14.47& 1.98& 26.97& 7637 (3076)& 2324 (774)\\
6 & 2020-02-17\,08:42:51 & 70.8 & 2.54& 23.27& 2.21& 22.60& 18083 (12426)& 3569 (2525)\\
\hline
\end{tabular}}}
\hspace*{5mm}}
\begin{flushleft}
\footnotesize{
\textbf{Notes.} For the exposure time we list both the total exposure time as well as the net exposure time of the stacked images (between brackets), for which images significantly contaminated by background stars were excluded.}
\end{flushleft}
\end{table}

\section{Analysis and Results}
To increase the signal-to-noise ratio of our images, we stacked the individual exposures within each of the six visits. We first identified stars that coincided with Borisov's extended coma using archival Digitized Sky Survey images\footnote{The STScI Digitized Sky Survey: \url{http://archive.stsci.edu/cgi-bin/dss_form}} of the same part of the sky. We discarded exposures with stars in the central 20-pixel aperture or with extremely bright stars in the central 50-pixel aperture. In the fifth visit the comet passed a crowded star field and to keep enough exposures we relaxed this restriction and only removed exposures with stars in the central 10-pixel aperture. The remaining exposures were then aligned and co-added using the astrometric position on the CCD from JPL/Horizons. 

The final (sixth) visit consists of two time longer total exposure time and co-adding would produce a crowded stacked background in the stack, contaminating the comet. Therefore, before co-adding, for every remaining February exposure we first clipped all pixels whose brightness exceed the peak of the comet to remove background sources, and filled every pixel with the azimuthally median value of pixels at the same comet-pixel distance as the filled pixel, and finally stacked the exposures. Because of the long time of exposures and movement of background sources, most clipping patterns such as star edges were also well suppressed during co-adding, which well addressed the crowded background problem.

\begin{figure}[t]
\centering
\includegraphics[scale=0.6]{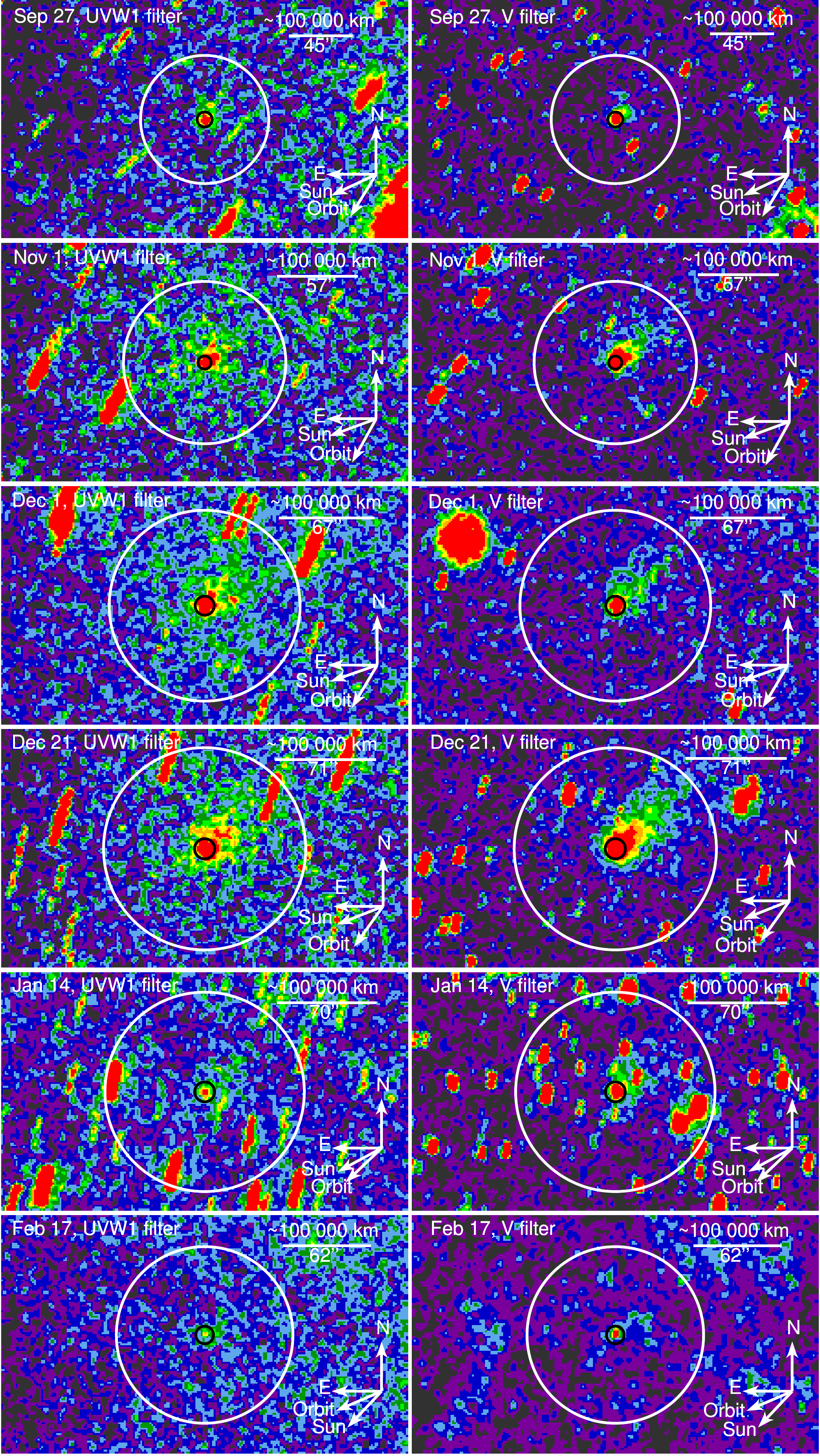}
\caption{Co-added \emph{Swift}/UVOT observations of 2I/Borisov (Left: UVW1; Right: V). All circles are centered on positions of the comet nucleus. White circles indicate photometric apertures used to measure water production rate and all have radii of 100\,000 km, with lengths of the scale indicators equaling to these radii. Black circles indicate apertures used to measure $A(\theta)f\rho$ and $A(0)f\rho$, with radii of 10\,000 km for the next five visits and a larger radius of 12\,000 km for the September visit to ensure the aperture exceeded UVOT's PSF. All panels have the same physical scale (170$\times$285 arcsec), and have been individually stretched linearly for optimal presentation, with north up, east to the left.} \label{fig:morphology} 

\end{figure}

The comet was clearly detected in the co-added V-band images of every visit, with a tail towards the anti-solar direction (Fig.~\ref{fig:morphology}). In the UVW1 images of middle four visits, this tail is mostly absent, and an extended coma within a radius of $\approx$100\,000~km can be clearly seen; at larger distances, background variations obscure the comet. In the first and final UVW1 image, the comet almost disappeared.

\subsection{Dust content ($A(\theta)f\rho$)}
 The V-band mostly samples solar continuum reflected by the dust in the coma. While the V filter includes the emission features of the $\Delta\nu=0,-1$ of the C$_2$ Swan-band sequence, several observers reported that Borisov was initially depleted in C$_2$  \citep{Kar19, Opi19}, implying that the contribution of C$_2$ emission to fluxes measured in the V-band is negligible. \citet{ban20} reported that C$_2$ production started in earnest after mid-October, and Borisov has become moderately depleted after November 26, we will estimate the effect in Section \ref{subsec: uncertainty}.
 
 We used the stacked images to measure the comet's V-band magnitudes and to derive $A(\theta)f\rho$, a measure of the dust content in the coma \citep{AHe84}.  To determine $A(\theta)f\rho$, smaller apertures are more desirable because these include less emission from gas. We used circular apertures of a fixed radius of 10\,000~km centered on the nucleus for the visits between November and February (corresponding to 5.7, 6.7, 7.1, 7.0 and 6.2~pix) and a slightly larger aperture of 12\,000~km (5.3 pix) for the September visit to ensure it was larger than UVOT's point spread function (5 pix) and the comet's apparent motion (3.4 -- 7.0~pix/exposure).
 
 Background regions were selected from nearby parts of the detector that have comparable systematic noise and avoided the extended coma and field stars.

We calculated magnitudes $m$ for the V and UVW1 band using the relation $m=Z_\mathrm{pt}-2.5\mathrm{log}(C)$, where $Z_\mathrm{V} = 17.89$~mag and $Z_\mathrm{UVW1} = 17.49$~mag are the photometric zero-points of the filters \citep{Poo08}, and $C$ is the count rate. $A(\theta)f\rho$ can then be determined by (derived from \citealt{AHe84}):
\begin{equation*}
A(\theta)f\rho=\frac{(2\Delta r_\mathrm{h})^2}{\rho}10^{0.4(m_{\sun}-m_\mathrm{V})}
\end{equation*}
where $\rho$ is the radius of the photometric aperture and $m_{V}$ is the measured magnitude of the comet. For the the solar magnitude at 1 AU through the same filter, we used $m_{\sun}=-26.75$~mag, which we estimated by convolving a solar spectrum model\footnote{Solar spectrum from \citet{col96}: \url{http://wwwuser.oats.inaf.it/castelli/sun/sun_ref_colina96.asc}} \citep{col96} with the V filter's effective area. Finally, we normalized $A(\theta)f\rho$ to a phase angle of 0~deg with the empirical phase function from D.~Schleicher\footnote{Composite dust phase function for comets by D.~Schleicher: \url{https://asteroid.lowell.edu/comet/dustphase.html}}. The resulting values for $A(0)f\rho$ are listed in Table~\ref{tab:results}. 

\subsection{Water Production Rates} \label{subsec:q_result} 
OH is produced by photolysis of H$_2$O in the coma and is commonly used as a direct proxy of comets' water production rates (cf. \citealt{Ahe95}). 
The UVW1 filter is well-placed to map fluorescent emission from the OH $A^2~\Sigma^+-X^2~\Pi$ band between 280 -- 330 nm \citep{Bod14}.  
To determine the flux of OH, we need to remove the continuum contribution to the UVW1 filter. For this, we developed an iterative method, where we first assume that the dust is a grey reflector and then adjust the dust reddening until modeled OH column density profiles best reproduce the observed profiles. 

To obtain `pure' OH images $C_\mathrm{OH}$ we subtracted  the  co-added image V-filter $C_\mathrm{V}$ from the co-added UVW1-filter image $C_\mathrm{UVW1}$ for every visit: $C_\mathrm{OH}=C_\mathrm{UVW1}-\alpha_0\cdot C_\mathrm{V}$, all in units of count rates, assuming contribution from C$_2$ emission is negligible. The removal factor $\alpha$ is the ratio of continuum count rates as measured with the two filters ($\alpha_0=0.093$ for a solar spectrum).  
Gas production rates are best extracted from larger apertures, where the contribution of the dust to the UVW1 filter is reduced owing to the different distribution of gas and dust in the coma. We masked all identifiable stars within an extraction region of 100\,000~km (45 -- 71 pixels) and all pixels around the center of the nucleus within a radius of 6 pixels to avoid smearing. We then derived median surface brightness profiles from the remaining pixels. To do that, we converted the net count rate of OH to the units of flux using:
\begin{math}
F_\mathrm{OH}=\beta \cdot C_\mathrm{OH}
\end{math}.
To estimate the factor $\beta$, which convert counts into units of flux, we used an OH spectral model from \citet{Bod19} and convolved this with the effective area of the UVW1 filter. This yields $\beta = 1.275\times10^{-12}\, \mathrm{erg}\,\mathrm{cm}^{-2}\,\mathrm{cts}^{-1}$.  Next, we converted the surface brightness profiles into OH column density profiles. To this, we used heliocentric velocity-dependent fluorescence efficiencies for the 1-0, 0-0 and 1-1 transitions in the OH $A^2~\Sigma^+-X^2~\Pi$ band \citep{Sch88}, scaled with the inverse square of the heliocentric distance, $r_\mathrm{h}^{-2}$. The total number of OH radicals in the aperture can be obtained by integrating the column density profiles over all annuli . We determined water production rates by comparing these measured OH contents with calculations from the vectorial model\footnote{Web Vectorial Model: \url{https://www.boulder.swri.edu/wvm/}}, which  describes the density distribution of neutral molecules in cometary atmospheres \citep{Fes81}. For this,  we assumed lifetimes of $8.6\times10^4$ s for $\mathrm{H_2O}$ and $1.29\times10^5$ s for OH, which are appropriate for the current solar minimum (both at 1~AU, scaled by $r_\mathrm{h}^{-2}$; \citealt{Hue92, Com04}), and we assumed a bulk H$_2$O outflow velocity of $0.85\times r_\mathrm{h}^{-0.5}\,\mathrm{km\,s^{-1}}$, and a constant OH velocity of 1.05 $\mathrm{km\,s^{-1}}$ \citep{Com04}.

\begin{figure}[t]
\centering
\includegraphics[scale=.57]{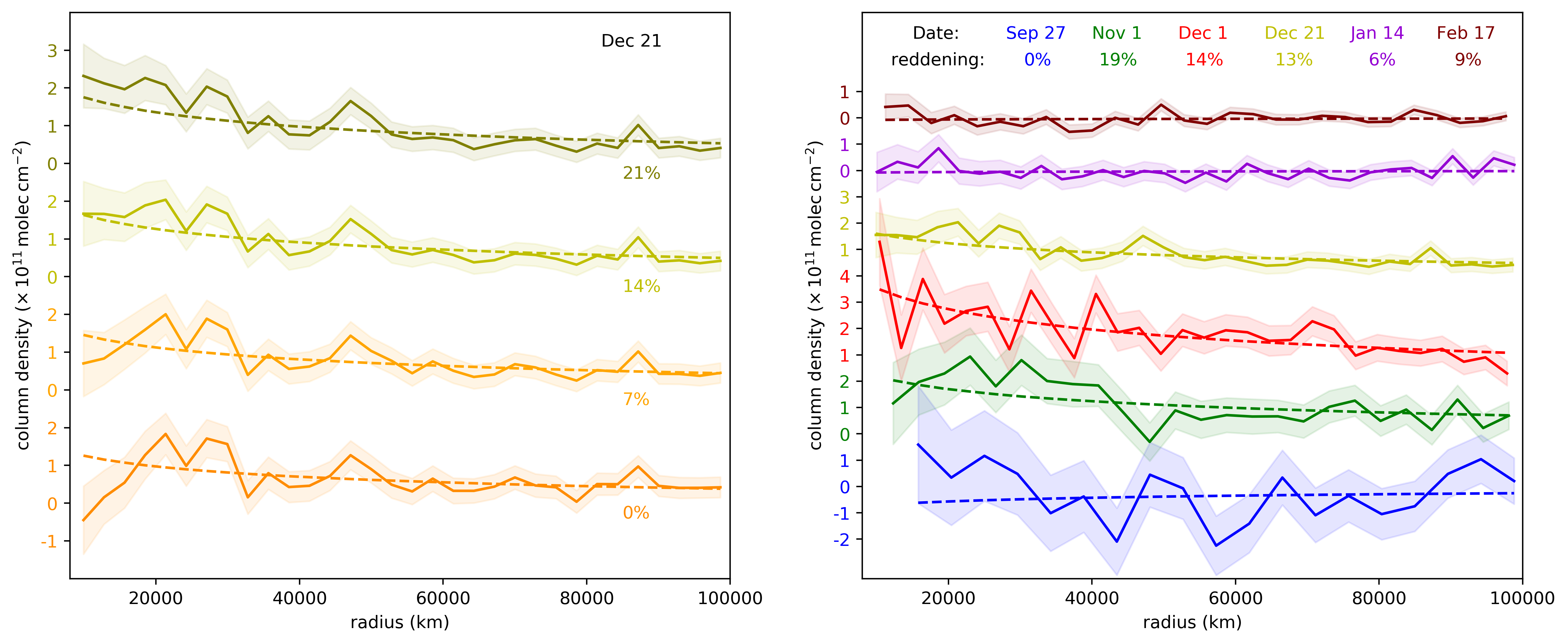}
\caption{{\bf Left:} Column density profiles with different continuum colors for the visit on Dec 21. Solid curves indicate column density profiles produced by removing scaled V-band images from observed UVW1-band images, adjusted for different levels of continuum reddening (\% per 100~nm between between 260 and 547 nm). Shaded areas indicate 1-$\sigma$ stochastic errors. Dashed curves are vectorial model profiles scaled to match the measured curves. {\bf Right:} Comparison of best fitted column density profiles for the 6 visits, with optimal reddening marked. Same symbols are in the left panel.}

\label{fig:profiles} 
\end{figure}

 At the same time, a modeled column density profile can be derived by scaling the vectorial model with the water production rate for every visit. Except for the Sep.~27, Jan.~14 and Feb.~17 visits, we noted that the modeled profiles could reasonably explain the observed column density profiles, but over-subtracted the continuum near the nucleus, resulting in negative values (Fig.~\ref{fig:profiles}). Given the large separation in wavelength of the UVW1 and V-band filter and our initial assumption of grey dust, this is likely due to the reddening of the dust. To address this, we empirically adjusted the reddening by changing the continuum removal factor $\alpha$, and repeated the steps above to derive new production rates and thus new modeled distribution, so that the measured column density profiles best matched the vectorial model distribution. For that, we assumed that the water production does not vary significantly on the timescale of a day, that most water is produced by the nucleus, and that there is no significant color gradient in the coma. We considered colors between 0 and 25~$\%$ per 100~nm between 260 and 547~nm with a step of 1~$\%$, and used a least-squares method to determine which dust color resulted in the best agreement between the modeled OH distribution and the \emph{Swift} observations.

As can be seen in  Fig.~\ref{fig:profiles}, the UVW1 image from the Sep. 27 visit has poor SNR and the resulting surface brightness profile does not match the OH models. This leads us to conclude that we did not detect OH. For the visits from Nov. 1 to Dec. 21, we find colors of $=19\%, 14\%\,\mathrm{and}\,13\%$ per 100 nm between 260 and 547~nm ($\alpha = 0.054, 0.062, 0.064$). These are consistent with most published results for the same wavelength range \citep{Fit19,Lin19, Guz19}, but not consistent with $-8\pm7\%$ per 100~nm reported by \citet{Kar19}. If the dust color is assumed to be constant during the short observation campaign, reddening can be considered as its mean value, 15\%.  The flat profiles of the final two visits lead us to conclude that we did not detect OH in those visits. Though their reddening are 6\% and 9\%, diverging from the average value 15\%, considering the non-detection of OH there will be large uncertainties of reddening based purely on vectorial model fit. Both reddening and water production rates results are summarized in Table~\ref{tab:results}.

\begin{table}[h]
\caption{Summary of results.}\label{tab:results}
\centering
\setlength{\tabcolsep}{6pt}
\scalebox{0.7}{
\hspace*{-5mm}
\makebox[\textwidth]{%
 \tabcolsep=0.1cm
 \begin{tabular}{*{15}{c}}
    \hline
    \hline
  \#&Midtime&  $r$FoV& Filter&  $m_\mathrm{filter}$& $F_\mathrm{filter}$$^1$ & $F_\mathrm{OH}$& Red. & \emph{g}(OH)$^2$ & $N_\mathrm{mol}$& $Q_\mathrm{H_2O}$& $A_\mathrm{act}$& min. r&  $A(\theta)f\rho$& $A(0)f\rho$ \\
    &(UT)&(arcsec/$\times10^4$ km)& &  (mag)& ($10^{-12}$ erg\,s$^{-1}\,\mathrm{cm}^{-2}$)& ($10^{-12}$ erg\,s$^{-1}\,\mathrm{cm}^{-2}$)& (\%)& (erg\,s$^{-1}\,\mathrm{molec}^{-1}$)& ($10^{31}$)& ($10^{26}$\,s$^{-1}$)& (km$^{2}$)& (km)& (cm)& (cm) \\
    \hline
    1&2019-09-27\,08:52:40& 5.3/1.2 ($A(\theta)f\rho$)& V& 18.1\textpm0.06& 0.2\textpm0.01&  $<\,0.5$& 0& $3.7\times10^{-16}$& $<\,3.4$& $<\,8.2$& $<\,2.9$& $<\,0.48$& 54.5\textpm3.1& 99.0\textpm5.6 \\
    && 45/10.0 ($Q_\mathrm{H_2O}$)& UVW1& $>\,18.6$& --& & & & & & & & & \\
    2&2019-11-01\,19:52:26& 5.7/1.0 ($A(\theta)f\rho$)& V& 17.5\textpm0.04& 0.3\textpm0.01& 1.0\textpm0.2& 19& $5.5\times10^{-16}$& 3.1\textpm0.7& 7.0\textpm1.5& 1.4\textpm0.3& 0.33\textpm0.04& 48.9\textpm1.9& 105.6\textpm4.1 \\
    && 57/10.0 ($Q_\mathrm{H_2O}$)& UVW1& 17.2\textpm0.1& 4.0\textpm0.5& & & & & & & & & \\
    3&2019-12-01\,12:17:04& 6.7/1.0 ($A(\theta)f\rho$)& V&  17.1\textpm0.06& 0.5\textpm0.03& 2.2\textpm0.2& 14& $5.5\times10^{-16}$& 4.8\textpm0.5& 10.7\textpm1.2& 1.7\textpm0.2& 0.37\textpm0.02& 43.4\textpm2.5& 101.4\textpm5.8 \\
    && 67/10.0 ($Q_\mathrm{H_2O}$)& UVW1& 16.7\textpm0.1& 3.5\textpm1.0& & & & & & & & & \\
    4&2019-12-21\,12:52:31& 7.1/1.0 ($A(\theta)f\rho$)& V& 17.1\textpm0.03& 0.5\textpm0.01& 1.6\textpm0.3& 13& $7.7\times10^{-16}$& 2.2\textpm0.4& 4.9\textpm0.9& 0.8\textpm0.1& 0.25\textpm0.02& 38.4\textpm1.0& 90.3\textpm2.5 \\
    && 71/10.0 ($Q_\mathrm{H_2O}$)& UVW1&  16.8\textpm0.1& 4.7\textpm0.6& & & & & & & & & \\
    5&2020-01-14\,10:36:11& 7.0/1.0 ($A(\theta)f\rho$)& V& 17.4\textpm0.05& 0.4\textpm0.02& $<\,3.1$& 6& $13.8\times10^{-16}$& $<\,2.7$& $<\,6.2$& $<\,1.2$& $<\,0.31$& 35.1\textpm1.6& 80.3\textpm3.7 \\
    && 70/10.0 ($Q_\mathrm{H_2O}$)& UVW1&  17.5\textpm0.66& --& & & & & & & & & \\
    6&2020-02-17\,08:42:51& 6.2/1.0 ($A(\theta)f\rho$)& V& 18.3\textpm0.06& 0.17\textpm0.009& $<\,0.7$& 9& $11.4\times10^{-16}$& $<\,0.9$& $<\,2.3$& $<\,0.8$& $<\,0.25$& 27.7\textpm1.5& 57.5\textpm3.1 \\
    && 62/10.0 ($Q_\mathrm{H_2O}$)& UVW1&  19.3\textpm0.85& --& & & & & & & & & \\
    \hline 
 \end{tabular}}
 \hspace*{5mm}
}

\begin{flushleft}
\footnotesize{
\textbf{Notes.} All errors are 1-$\sigma$ stochastic errors, except the uncertainty of the magnitudes which include photometric zero-points errors. Upper or lower limits are 3-$\sigma$ stochastic errors, based on count rates measured in UVW1 with no continuum removed. V-band and $A(0)f\rho$ measurements were measured from an aperture with radius $r$FoV ($A(\theta)f\rho$) in this table. UVW1-band measurements and other quantities were derived from $r$FoV ($Q_\mathrm{H_2O}$). 

$^1$ No OH gas emission flux is included in $F_\mathrm{UVW1}$. 

$^2$ We acquired \emph{g}(OH) from \citet{Sch88}, accounting for the comet's heliocentric velocity and scaled by $r^{-2}_\mathrm{h}$.

}
\end{flushleft}
\end{table}

\subsection{Uncertainties} \label{subsec: uncertainty}
The results are subject to several uncertainties. Uncertainties in the calibration introduce systematic errors. \emph{Swift}/UVOT is very well calibrated with an accuracy better than 4\% \citep{Poo08}, but its sensitivity has reportedly degraded over time by about 1\% per year and is wavelength dependent \citep{bre11}. The resulting fluxes may thus be underestimated, leading to lower measured values of $A(0)f\rho$. A 10-percent decrease of effective area can introduce 11\% of underestimation of water production rate. UVOT suffers from coincidence losses at high photon flux ($>10$ counts s$^{-1}$, \citealt{Poo08}), which does not apply to Borisov, whose maximum count rate was about 7~counts~s$^{-1}$. The systematic uncertainty in our water production rate is most likely driven by the models used, including uncertainties in the \emph{g}-factors, velocities, and lifetimes of parent and fragment species. 

Relative errors from photon counting, background subtraction and continuum subtraction are propagated to calculate stochastic uncertainty. No OH is detected in the September, January, and February visits (Section \ref{subsec:q_result}). For the other three visits the resulting stochastic errors in the continuum removed OH images are 21\%, 11\% and 19\%, which account for a major portion of the total uncertainties. For the six visits the V-band stochastic errors are small, at 5\%, 4\%, 6\%, 2\%, 4\% and 5\% respectively.

\emph{Swift} did not track Borisov, which introduced some smearing. We evaluated its effect on our measurement of $A(0)f\rho$ by modeling the dust distribution assuming a simple Gaussian model. We then co-added 50 evenly distributed Gaussian curves across the distance of smearing during a single exposure. We found that the relative difference caused by the apparent motion on our photometry was less than 3\% when using apertures with radii larger than 10\,000 km.

The broadband UVW1 filter not only covers OH $A^2~\Sigma^+-X^2~\Pi$ fluorescent emission, but also includes emission features of CS, NH and CN. The limited transmission of the UVW1 filter at the wavelength of these features implies that their count rate contributions are typically more than an order of magnitude fainter than the OH lines \citep{Bod14}. The V-band is contaminated by the $\Delta\nu=0,-1$ Swan-band sequence of C$_2$ molecules. This C$_2$ emission will lead to an oversubstraction of the continuum from the UVW1 flux and will result in an underestimation of $Q_\mathrm{H_2O}$. To evaluate this, we assumed C$_2$ abundances relative to OH to be less than 1/1000 for all visits \citet{Ahe95}. The ratio is consistent with derived ratios of $Q_\mathrm{C_2}$ detected by \citealt{Lin19} and \citealt{ban20} to our measured $Q_\mathrm{OH}$ (methods to determine $Q_\mathrm{OH}$ are discussed in Section \ref{subsec:q_discuss}). At the band heads of $\Delta\nu=0$ and $\Delta\nu=-1$, UVOT's effective area is about 10~cm$^2$ and 20~cm$^2$. Using fluorescence efficiencies from \citet{AHe82}, we estimate that C$_2$ leads to underestimating of $Q_\mathrm{H_2O}$ by less than 15\%.

To estimate uncertainties in the OH flux introduced by the dust color, we compared our optimized water production rates with those derived by assuming reddening from 0\% to 25\%. That indicates the largest differences of 44\%, 7\% 22\% for the production rates measured in the November and two December visits. 

\section{Discussion} \label{sec:discussion}
The direct detection of OH emission by \emph{Swift}-UVOT and the resulting characterization of Q$_{H2O}$ allows us to compare to previously published results, investigate the minimum active area required, and attempt to place 2I/Borisov's chemistry and activity, as it is currently understood,  in the context of previously observed Solar System comets. Additionally, the availability of visible imaging taken near simultaneously allows a dust to gas ratio to be estimated. Combining these results from \emph{Swift} with previously reported results provides a more complete picture of Borisov and perhaps the system it was ejected from. 
\begin{figure}[h]
\centering
\includegraphics[scale=1.0]{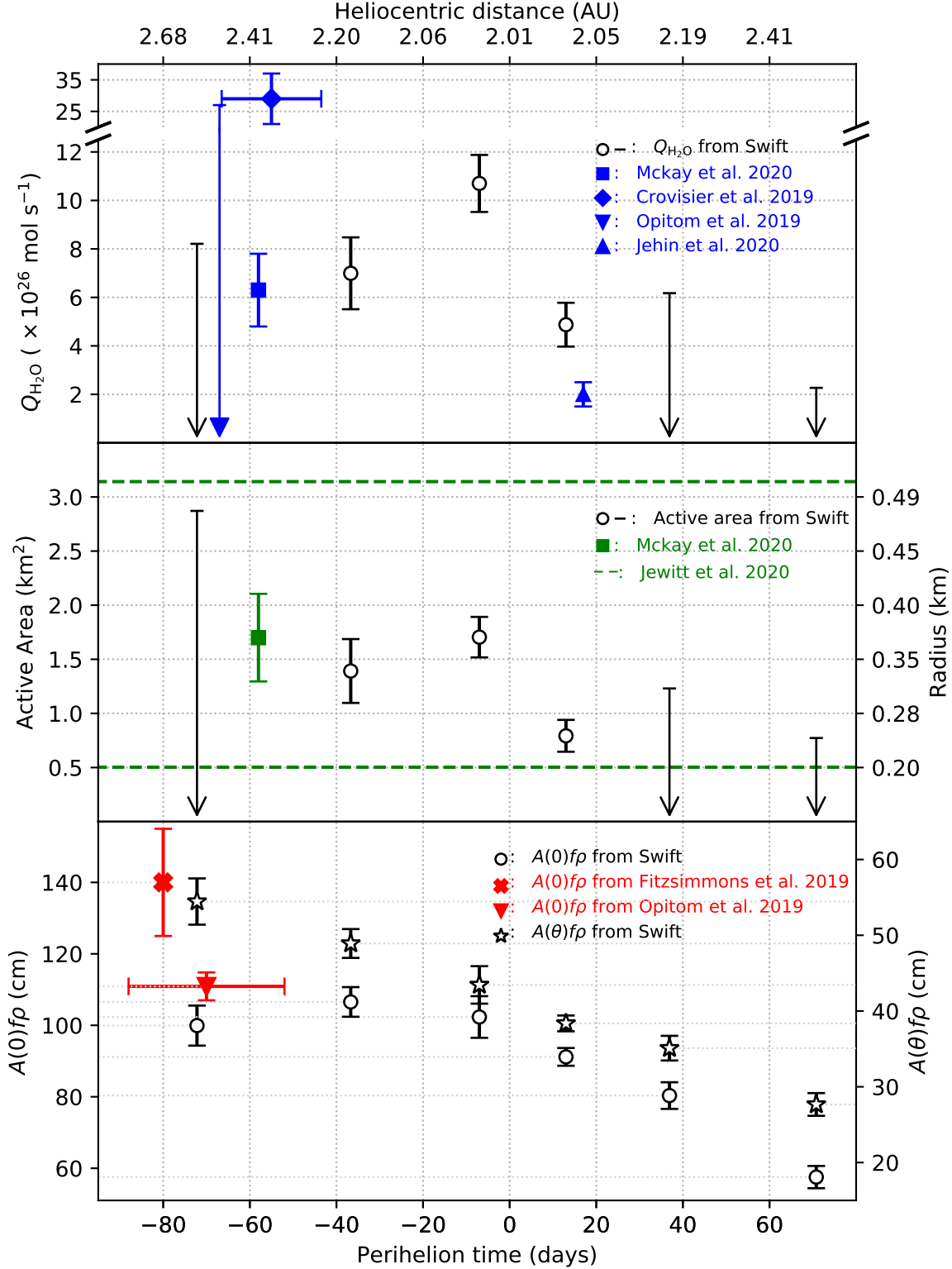}
\caption{Activity of 2I/Borisov over time. Results of water production rates (blue, top panel), minimum active area and converted minimum radius (green, middle panel), $A(0)f\rho$ and $A(\theta)f\rho$ (red, bottom panel) are marked by different colors. Results from different instruments are distinguished by markers' shapes: \emph{Swift}/UVOT (this work, open circles, open stars and the upper limits marked by simple arrows); APO/ARCES (\citealt{McK19}, squares); NRT (\citealt{Cro19}; diamonds); WHT/ISIS (\citealt{Opi19}, downward triangles and the upper limit marked by a filled arrow; UVES/VLT (\citealt{Jeh20}, an upward triangle); and TRAPPIST-North (\citealt{Fit19}, filled crosses). Values of $Q_\mathrm{OH}$ reported by \citet{Cro19} and \citet{Opi19} were converted into $Q_\mathrm{H_2O}$ by an empirical relation from \citet{Coc93}. The $A(0)f\rho$ by \citet{Opi19} was color-corrected from the R to the V-band (Section \ref{subsec:dust_disc}). We indicate the current narrowest constrains of the comet's radius from HST/WFC3 \citep{Jew19} by green dashed lines in the middle panel. Horizontal errorbars indicate time spans of observations.}
\label{fig:comparison} 
\end{figure}

\subsection{Water production rates and active area}\label{subsec:q_discuss}
We did not detect any OH on the first visit on Sep.~27, 2019 with a 3-$\sigma$ upper limit of $Q_\mathrm{H_2O}<8.2\times10^{26}$~molecules~s$^{-1}$. Between Nov.~1 and Dec.~1, the water production rate increased from $(7.0\pm1.5)\times10^{26}$ to $(10.7\pm1.2)\times10^{26}$ molecules~s$^{-1}$, and it appears to decrease rapidly after that, to $(4.9\pm0.5)\times10^{26}$~molecules~s$^{-1}$ on Dec. 21. The OH coma almost disappeared in the final two visits, with a 3-$\sigma$ upper limit of $Q_\mathrm{H_2O}<6.2\times10^{26}$ and $2.3\times10^{26}$~molecules~s$^{-1}$.
These results are in good agreement with those acquired by others (Fig.~\ref{fig:comparison}). \citet{McK19} used spectroscopy to detect the emission of [OI] 630 nm and derived a water production rate of $(6.3\pm1.5)\times10^{26}$ molecules~s$^{-1}$ on UT Oct.~11, 2019, when Borisov was at 2.38~AU from the Sun. Our results confirm that most of the [OI] emission is indeed likely the product of emissive photodissociation of H$_2$O (as opposed to different physical processes and/or parent species, cf. \citealt{Bod16}). \citet{Cro19} used the Nancay radio telescope between Oct. 2 -- 25 and reported a tentative OH production rate of $(3.3\pm0.9)\times10^{27}$ molecules~s$^{-1}$. We converted OH production rates from \citet{Cro19} using the empirical conversion formula from \citet{Coc93}, $Q_\mathrm{H_2O}=1.361\,r_\mathrm{h}^{-0.5}\,Q_\mathrm{OH}$, which gives a water production rate of ($2.9\pm0.8)\times10^{27}$ molecules~s$^{-1}$. \citet{Opi19} used spectroscopic observations with the 4.2-m William Herschel Telescope to search for OH emission around 308~nm and reported a 3-$\sigma$ upper limit for the production rate of OH of $2\times10^{27}$ molecules~s$^{-1}$ on Oct.~2, 2019. To convert this $Q_\mathrm{OH}$ into $Q_\mathrm{H_2O}$, considering that \citet{Opi19} has included the heliocentric relation for the gas outflow velocity $v_\mathrm{gas}\approx r_\mathrm{h}^{-0.5}$, we used $Q_\mathrm{H_2O}=1.361\,Q_\mathrm{OH}$ and found $Q_\mathrm{H_2O}=2.7\times10^{27}$ molecules~s$^{-1}$. Finally, \citet{Jeh20} used UVES spectrograph on 8.2-m Very Large Telescope to measure OH emission lines at 310~nm, and obtained $Q_\mathrm{H_2O}=2\times10^{26}$ molecules~s$^{-1}$ between Dec.~24 and Dec.~26, 2019. $Q_\mathrm{H_2O}$ from \citet{Cro19} is larger than those found by us and the NUV/optical spectroscopic studies \citep{Opi19,McK19}. Part of this difference may be explained by assumptions regarding the heliocentric trend of the outflow velocity of water or other details of the modeling.

To further investigate the comet's evolving activity, we calculated the minimum active area corresponding to the water production rates using a sublimation model\footnote{NASA PDS Small Body Node Tools -- Sublimation of Ices: \url{https://pds-smallbodies.astro.umd.edu/tools/ma-evap/index.shtml}} \citep{Cow79}. Here, we assume that every surface element has constant solar elevation (as would be the case if the rotational pole were pointed at the Sun or if the nucleus was very slowly rotating) and is therefore in local, instantaneous equilibrium with sunlight. This maximizes the sublimation averaged over the entire surface, and results in a minimum total active area. We further assumed a Bond albedo of 0.04 and an infrared emissivity of 100\%. The resulting minimum active areas $A_\mathrm{min}$ are shown in Fig.~\ref{fig:comparison}. Our results indicate that when Borisov approached the Sun, the minimum active area remained approximately constant with a slight increase from 1.4 to 1.7 km$^2$, but then decreased dramatically to less than half of that to 0.8~km$^2$ immediately after perihelion. We note that the limits on the active area for the Sep.~27, Jan.~14 and Feb.~17 measurement has no physical meaning (it is an upper limit for a minimum active area), but it demonstrates that \emph{Swift}'s non-detection of OH is consistent with the comet's activity levels in the following epochs. 

Converting these areas into minimum radii by $r_\mathrm{min} = \sqrt{A_\mathrm{min}/4\pi}$, we find corresponding minimal radii of 0.33, 0.37, and 0.25~km, respectively. Currently, the tightest constraints of the nucleus' size were achieved by combining a model of the non-gravitational forces on the nucleus with observations by the Wide Field Camera 3 on the Hubble Space Telescope, which yielded 0.2 km $<$ r $<$ 0.5 km \citet{Jew19}. The minimum radii derived from our active area estimates fall well within this range (Fig.~\ref{fig:comparison}) and imply that a significant fraction ($>55$\%) of the surface of Borisov is active. This level is comparable to a small set of very active Jupiter-family comets (cf. \citealt{Com19}) and several dynamically new and young comets from our solar system \citep{Bod14,Bod15}. 

To assess the total mass loss pre-perihelion, we assumed the water production rate was negligible outside 4.0~au. We then integrated the linearly interpolate water production rates time until Dec.~25, when \citet{Jeh20} detected OH, and obtained a total loss of $6\times10^{33}$~molecules until that time, which corresponds to $2.3\times10^8$~kg of ice in nucleus. Assuming a gas mixing ratio of 9\% CO and 7\% CO$_2$, the average values from \citet{boc04}, and all the other is H$_2$O, then the total mass of the volatiles lost are about $3.2\times10^8$~kg. Assuming a dust-to-gas ratio of 4 as observed by Rosetta around 67P/Churyumov-Gerasimenko \citep{Rot15}, the comet lost approximately $1.6\times10^9$~kg of material on its trajectory until Dec.~21. Assuming a spherical shape of the nucleus, the size range from \citet{Jew19} and a density of 500~kg\,m$^{-3}$, 0.6\% -- 9.4\% of the entire mass was lost, corresponding to a global layer of 1.0 -- 6.4~m . While this mass loss is an extremely crude estimate which depends on multiple assumptions, this number compares well to observed loss rates of comets 67P (0.1\% \citet{Pat19} and 103P/Hartley~2 ($>1\%$; \citet{Tho13}).

Based on the current constraints of the size of the nucleus and our derived minimum active area, we cannot exclude that the activity levels require an active area larger than the nucleus' surface, possibly pointing to the presence of an additional source of H$_2$O in the coma such as icy grains \citep{AHe84,AHe11,Pro2018}. Based on Borisov's high levels of activity at large heliocentric distance, \cite{Sek19} inferred the presence of a halo of icy grains. The presence of icy grains is relevant because their physical properties may sample that of the primordial ice contained within the nucleus \citep{Pro2018} and can skew remote relative abundance measurements of the gases in the coma \citep{Bod14, Kel17}. The comparison between $Q_\mathrm{H_2O}$ measured by large apertures and narrow slits can be employed to assess whether there was an extended source. As introduced by \citet{Bod14}, in comet C/2009 P1 (Garradd), a dichotomy was observed between production rates based on spectra acquired with narrow slits (covering the inner 1\,000~km of the coma) compared to those derived from observations acquired with much larger apertures. This difference was attributed to the existence of an icy grain halo in the inner region. For Borisov, the water production rates derived from our large-aperture observations are in good agreement with those derived from narrow-slit observations (3.2 $\times$ 1.6 arcsec; apparent size $\approx$ 6\,500~km; \citet{McK19}. However, this does not allow us to conclusively exclude the presence of icy grains, because this slit size is much larger than the expected lifetimes of icy grains \citet{Yan19}. Icy grains would have sublimated within the slit used by McKay et al., which is thus too large to help us find different production rates with both methods. In addition, \citet{Yan19} used the GNIRS spectrograph at the 8-m Gemini telescope and found that large or pure ice grains, if present, comprise no more than 10\% of the coma cross-section within their (smaller) slit of 2\,830~km when Borisov was 2.6~AU from the Sun. Even if present, at that level icy grains would contribute little to Borisov's water production rate.

\subsection{Temporal evolution of production rates}
In order to compare the activity evolution of Borisov with that of Solar System comets, We used $Q_\mathrm{H_2O}$ derived from \citet{McK19} and our visits before perihelion to calculate the slope of $Q_\mathrm{H_2O}$ of $r_\mathrm{h}$ in a log-log plot and obtained the result of $-3$. \citet{Com2018} reported this slope for 11 dynamically new comets and 13 Jupiter-family comets (JFCs) in different apparitions. The pre-perihelion slope of -3 of 2I/Borisov is steeper than those of all dynamically new comets reported by \citet{Com2018}, ranging from $-3$ to $-0.8$, and is at the shallower end of the wide range of reported Jupiter-family comets, which is from $-14$ to $-0.4$ \citep{Com2018}. To compare our results with \citet{Ahe95}, we convert our water production rates into OH production rates (Section \ref{subsec:q_discuss}), and calculated a $Q_\mathrm{OH}$ power-law exponent of $-2.5$. Compared to \citet{Ahe95}, Borisov's slope is again steeper than that of the only reported dynamically new comet (C/1980 E1 Bowell), $-0.63$, and shallower than three of the four reported JFCs ranging from $-7.91$ to $-1.78$. 

It is of note that  the $Q_\mathrm{H_2O}$ pre-perihelion power-law exponent of Borisov, $-3$, was similar to that of two dynamically new comets; C/2009 P1 (Garradd) and C/2013 A1 (Siding Spring). Both of these comets were observed to have power-law exponents of $-2.6$ \citep{Bod14,Bod15}. In both cases the steep increases were both attributed to onset of sublimation of icy grains in coma. As discussed in \ref{subsec:q_discuss}, icy grains do not contribute significantly to the total water production rate, suggesting that the observed behavior is caused by seasonal or evolutionary processes.

After perihelion, $Q_\mathrm{H_2O}$ decreased with a slope steeper than -116.7, assuming a constant slope. That is much more rapid than all previously detected comets reported by \citet{Com2018}, whose steepest slope is -19.6. The non-detection of OH in January and February enhances the reliability of this rapid disappearance. More evidences are needed to determine the reasons of this rapid decrease, including surface erosion, nucleus rotation and even fragmentation \citep{mak01}.

\subsection{Relative abundance of fragment species} \label{subsec:abundance}
Our observations allow us to compare measured abundances of other fragment species (CN, C$_2$, C$_3$, and NH) with respect to the water production rates, and thus to compare the abundances of Borisov with those of solar system comets. \citet{Ahe95} provides the largest survey for fragment species abundances compared to OH. We obtained $Q_\mathrm{OH}$ from $Q_\mathrm{H_2O}$ as discussed in section \ref{subsec:q_discuss}. Our observations of Nov.~1 coincide with those reported by \citet{Lin19}, who observed Borisov between Nov.~1--5. This yields logarithmic ratios of $Q_\mathrm{X}/Q_\mathrm{OH}$ of $-2.4$, $-3.0$, and $-4.2$ for CN, C$_2$, and C$_3$. \citet{ban20} observed Borisov on Nov.~26, which is nearest to our visit on Dec.~1, and their results yield in logarithmic ratios of $Q_\mathrm{X}/Q_\mathrm{OH}$ of $-2.8$ and $-3.0$ for CN and C$_2$. These place Borisov solidly in the category of carbon-to-water depleted comets, for which \citet{Ahe95} reported mean values of CN/OH = $-2.7$, C$_2$/OH = $-3.3$, and C$_3$/OH = $-4.2$. Assuming a constant active area of 1.7~km$^2$, we used the sublimation model to estimate that around Sep.~27, the water production rate was $5.1\times10^{26}$ molecules~s$^{-1}$, well below our detection limit. This would yield a relative abundances of $Q_\mathrm{CN}/Q_\mathrm{OH}$ = $-2.2$ to $-2.1$ and $Q_\mathrm{C_2}/Q_\mathrm{OH}<-3.6$ using the production rates reported for Sep.~20 \citep{Fit19} and Oct.~1 \citep{Kar19}, again consistent with carbon-to-water depleted comets \citep{Opi19}.

\citet{ban20} also reported $Q_\mathrm{NH_2}$, which results in log$_{10}(Q_\mathrm{NH_2}/Q_\mathrm{OH})=-2.3$. \citet{Ahe95} notes the lack of distinction between typical and depleted comets based on their NH/OH but does not give statistic results of NH$_2$/OH. Therefore we calculated $Q_\mathrm{NH_2}/Q_\mathrm{H_2O}$ and find $4.49\times10^{-3}$, which exceeds the maximum value of $4.36\times10^{-3}$ for 50 well-observed comets reported by \citet{Fin09}. Despite the lack of adequate interstellar comparisons, this strongly indicates that Borisov is enriched in NH$_2$, consistent with deduction from \citet{ban20}.

\subsection{Gas to dust ratio}
\label{subsec:dust_disc}
Results of $A(\theta)f\rho$ and $A(0)f\rho$ are shown in Table~\ref{tab:results} and Fig.~\ref{fig:comparison}. Our values of $A(0)f\rho$ measured from V-band images range from 57.5~cm to 105.6~cm, with a peak before perihelion, while values of $A(\theta)f\rho$ decrease linearly with time. Our values of $A(0)f\rho$ are lower than results from observations with the $BVR_cI_c$ filter system of TRAPPIST-North, which include V-band $A(0)f\rho$ of $140\pm15$~cm on Sep.~20 from \citet{Fit19}, and R-band $A(0)f\rho$ of $132.4\pm4.7$cm between Sep. 11 and Oct. 17 from \citet{Opi19}. Like ours, these measurements used a 10\,000\,km-radius apertures around the nucleus, except for the 12\,000\,km-radius aperture used in our September visit to avoid PSF problems, and all are corrected by phase function from D. Schleicher. As mentioned in section \ref{subsec: uncertainty}, we consider that these difference in $A(0)f\rho$ partly arise from decline of effective area of UVOT as well as differences between Johnson and UVOT optical responses. For the latter factor, since the cometary dust is reddened (Section \ref{subsec:q_result}), measurements of $A(0)f\rho$ will vary for different filters. For a solar spectrum with reddening of 15\%, we determined that a color correction required to compare measurements of $A(0)f\rho$ with the V filter of \emph{Swift}/UVOT are $+1$\%  and $-16$\% for the V filter and R filter of Johnson system, respectively. These uncertainties are not adequate to explain the differences, which indicates that other as of yet unknown differences may be responsible for the discrepancies.

The logarithmic ratio between $A(0)f\rho$ and $Q_\mathrm{OH}$ remains stable at $-24.9$ from November to December. However, comparing to other comets is complicated, as we note that \citep{Ahe95} did not correct for phase effects. Therefore, we used $A(\theta)f\rho$ to recalculate dust-gas ratio and got results between $-25.4$ and $-25.1$, which are consistent with values of carbon-chain depleted solar system comets ($-25.94$~$<$ $A(\theta)f\rho$/OH $ < $~$-24.85$, \citealt{Ahe95}). We note that \cite{Ahe95} used a green filter (484.5 or 524~nm) for their $A(0)f\rho$, while we use the V filter (546.8~nm, FWHM 76.9~nm), this introduces around 6\% overestimation of $A(0)f\rho$ and less than 1\% overestimation of $A(\theta)f\rho$/OH for a solar spectrum with reddening of 15\%, that has negligible effects for our conclusion.

\section{Summary} \label{sec:summary}
2I/Borisov is the first notably active interstellar comet. The relatively early discovery, relative brightness, and placement in the sky made it possible to characterize its activity evolution, constrain the size and rotation of its nucleus, and for the first time conduct a chemical inventory of an extra-solar small body. We obtained optical and ultraviolet observations of the interstellar comet 2I/Borisov using \emph{Swift}/UVOT at heliocentric distances from 2.56~au pre-perihelion to 2.54~au post-perihelion. \emph{Swift} detected OH $A^2~\Sigma^+-X^2~\Pi$ emission in three of the four epochs, and we used these observations to derive water production rates, the corresponding minimum active areas, exponential power-law variations with heliocentric distances, relative abundances and gas-to-dust ratios. Our findings are: 
\begin{enumerate}
    \item Water production rates increased gradually between 2.56~au to 2.01~au from $(7.0\pm1.5)\times10^{26}$ to  $(10.7\pm1.2)\times10^{26}$ molecules~s$^{-1}$, and then decreased rapidly to $(4.9\pm0.9)\times10^{26}$ molecules~s$^{-1}$ at 2.03~AU post-perihelion.

    \item Using a sublimation model for the nucleus, we constrained the minimum active area of Borisov to 1.7~km$^2$. This corresponds to a minimum radius of 0.37~km. Comparing this to other published estimates of the size of Borisov's nucleus \citep{Jew19}, it is likely that at least 55\% of the surface is active. Icy grains do not contribute significantly to the bulk water production rates.
    
    \item Compared with broad comet surveys, the measured slopes of the water production rate with respect to heliocentric distance indicate that before perihelion, $Q_\mathrm{H_2O}$ ($Q_\mathrm{OH}$) of 2I/Borisov increases more steeply than most dynamically new comets, while the increase is at the slower end of the wide range of Jupiter-family comets. It should be noted that power-law exponents of the production rate trend with heliocentric distance of 2I/Borisov are similar to those of two dynamically new comets, C/2009 P1 (Garradd) and C/2003 A1 (Siding Spring), where the the sublimation of icy grains in coma likely contributed significantly to the total water production. After perihelion, 2I's $Q_\mathrm{H_2O}$ ($Q_\mathrm{OH}$) decreases much more rapidly than all previously observed comets reported by surveys.
    
    \item Our water production rates confirm that relative to water, 2I/Borisov is depleted of carbon-chain molecules and indicate that 2I/Borisov is enriched in NH$_2$.
    
    \item We find values of UVOT/V-band $A(0)f\rho$ varying between 57.5 and 105.6~cm with a slight trend peaking before its perihelion, while values of $A(\theta)f\rho$ decrease linearly with time. The dust-to-gas ratios from $A(\theta)f\rho$ are consistent with values of carbon-chain depleted solar system comets \citep{Ahe95}.
\end{enumerate}

We find that interstellar comet 2I/Borisov is in many regards similar to Solar System comets. Its specific properties (production rate slope, chemical composition, size estimate) do not firmly place it within any single one of the dynamical families. Its relatively low production rates made it a very challenging object to observe, and this may complicate placement into the Solar System comet taxonomy, which are generally biased towards brighter objects. 

\acknowledgments
We thank J. Gropp and the \emph{Swift} team for use of Director's Discretionary Time and for the careful and successful planning of these observations. This research has made use of NASA's Astrophysics Data System Bibliographic Services.

\vspace{5mm}
\facilities{\emph{Swift}(UVOT)}

\end{CJK*}
\bibliographystyle{aasjournal}
\bibliography{main}



\end{document}